\documentclass{aa}

\newcommand{\magpt}[2]{\mbox{$\rm #1\hspace{-0.25em}\stackrel{m}{.}
   \hspace{-1.0mm}#2$}}               % magnitude \magpt {}{}
\def\bsec{\hbox{$.\!\!{\arcsec}$}}

\newcommand\RA[3]{#1$^{\rm h}$#2$^{\rm m}$#3$^{\rm s}$}

\newcommand\DEC[4]{#1$^{\circ}$#2\arcmin#3\bsec#4}
\newcommand\teff{$ {\rm T_{eff}}$}

\newcommand\logg{$\log {\rm g}$}
\newcommand\loghe{${\rm \log{\frac{n_{He}}{n_{H}}}}$}
\newcommand\Ebv{$ {\rm E_{B-V}}$}
\newcommand\ebv{$ {\rm E_{B-V}}$}
\newcommand{\Lsolar}{\mbox{\,$\rm L_{\odot}$}}    % solar luminosity
\begin{document}
%\thesaurus{05 ()}
\title{Hot HB stars in globular clusters - Physical parameters and
consequences for theory. VI. The second parameter pair M~3 and M~13}
\author{S. Moehler\inst{1,2}\thanks{Based on observations 
collected at the German-Spanisch Astronomical
Center (DSAZ), Calar Alto, operated by the Max-Planck-Institut f\"ur
Astronomie Heidelberg jointly with the Spanish National Commission for
Astronomy}
 \and W.B.~Landsman\inst{3}
 \and A.V.~Sweigart\inst{4}
 \and F.~Grundahl\inst{5}}
\offprints{S. Moehler}
\institute{Institut f\"ur Theoretische Physik und Astrophysik, Abteilung
 Astrophysik, Leibnizstra\ss e 15, 24098 Kiel, Germany
(e-mail: moehler@astrophysik.uni-kiel.de)
 \and Dr. Remeis-Sternwarte, Astronomisches Institut der Universit\"at
Erlangen-N\"urnberg, Sternwartstr. 7, 96049 Bamberg, Germany 
 \and SSAI, NASA\,Goddard Space Flight Center, Code 681, Greenbelt,
 MD 20771, USA (e-mail: landsman@mpb.gsfc.nasa.gov)
 \and NASA\,Goddard Space Flight Center, Code 681, Greenbelt,
 MD 20771, USA (e-mail: sweigart@bach.gsfc.nasa.gov),
\and Institute of Physics and Astronomy, 
Aarhus University, Ny Munkegade, 8000 Aarhus C, Denmark
(e-mail: fgj@ifa.au.dk)
}
\titlerunning{Hot HB stars in M~3 and M~13}
\authorrunning{Moehler, Landsman, Sweigart, Grundahl}
%\date{Received, accepted }

\abstract{ We present the results of spectroscopic analyses of hot
horizontal branch (HB) stars in M~13 and M~3, which form a famous
``second parameter'' pair. From the spectra and Str\"omgren photometry
we derived -- for the first time in M~13 -- atmospheric parameters
(effective temperature and surface gravity).  For stars with
Str\"omgren temperatures between 10,000 and 12,000 K we found
excellent agreement between the atmospheric parameters derived from
Str\"omgren photometry and those derived from Balmer line profile
fits.  However, for cooler stars there is a disagreement in the
parameters derived by the two methods, for which we have no
satisfactory explanation. Stars hotter than 12,000~K show evidence for
helium depletion and iron enrichment, both in M~3 and M~13.
Accounting for the iron enrichment substantially improves the
agreement with canonical evolutionary models, although the derived
gravities and masses are still somewhat too low.  This remaining
discrepancy may be an indication that scaled-solar metal-rich model
atmospheres do not adequately represent the highly non-solar abundance
ratios found in blue HB stars affected by diffusion. We discuss the
effects of an enhancement in the envelope helium abundance on the
atmospheric parameters of the blue HB stars, as might be caused by
deep mixing on the red giant branch or primordial pollution from an
earlier generation of intermediate mass asymptotic giant branch stars.
\keywords{Stars: atmospheres -- Stars: evolution -- Stars: horizontal
branch -- Globular clusters: individual: M~3 -- Globular clusters:
individual: M~13 }}

\maketitle

\section{Introduction\label{sec_intro}} 

The two globular clusters M~3 and M~13 form a famous ``second
parameter'' pair of clusters that show very different horizontal
branch (HB) morphologies despite quite similar metallicities ([Fe/H] =
$-$1.57 for M~3 and $-$1.54 for M~13, Harris \cite{harr96}): M~3
possesses a horizontal HB populated from the red to the blue end,
while M~13 exhibits a predominantly blue HB followed by a very long
blue tail that extends to temperatures of $\approx$ 35,000~K (Parise et
al. \cite{pabo98}). No consensus has yet been reached on the reasons
for this difference in HB morphology. While it has often been argued
that age, one of the most widely discussed ${\rm 2^{nd}}$ parameter
candidates (Sarajedini et al.\ \cite{sach97}), cannot be the ${\rm
2^{nd}}$ parameter for this pair (Catelan \& de Freitas Pacheco
\cite{cade95}, Ferraro et al. \cite{fepa97}, Paltrinieri et
al. \cite{pafe98}, Grundahl \cite{grun99}), the subject remains
controversial (e.g.\ Rey et al. \cite{reyo01}). VandenBerg
(\cite{vand00}), in particular, has emphasized that something other
than age must be different between M~3 and M~13.

M~13 (along with $\omega$~Cen) harbors the most dramatic examples of
stars presenting abundance anomalies on the upper red giant branch
(RGB; Kraft et al. \cite{krsn93}, \cite{krsn95}, \cite{krsn97},
Cavallo \& Nagar \cite{cana00}), whereas the abundance anomalies in M~3 are
considerably less pronounced (Kraft et al. \cite{krsn92}, Cavallo \&
Nagar \cite{cana00}). In particular, the super-oxygen poor stars found
near the tip of the RGB in M~13 are absent in M~3. These abundance
anomalies may arise from deep mixing processes which bring up
nuclearly processed material from the vicinity of the hydrogen burning
shell during a star's RGB phase. Support for this possibility comes
from the larger Na and Al abundances and smaller O and Mg abundances
found in stars near the tip of the RGB in M~13 compared to stars
further down the RGB (Kraft \cite{kraf99}, Cavallo \& Nagar
\cite{cana00}). Alternatively, the recent detections of similar
abundance anomalies among main-sequence stars in a few clusters
(Cannon et al.\ \cite{cacr98}, Gratton et al.\ \cite{grbo01}) suggest
a primordial origin, perhaps due to pollution from an earlier
generation of intermediate mass asymptotic giant branch (AGB) stars.
       
Both the deep mixing and primordial scenarios for the origin of the
abundance anomalies may have consequences for the luminosity and
morphology of the HB, as outlined by Sweigart (\cite{swei97b}) and
D'Antona et al. (\cite{daca02}), respectively. Briefly, in the
deep-mixing scenario, helium from the H-burning shell is mixed
into the envelope on the RGB and thereby causes a star to have a
bluer and more luminous position on the HB. In the primordial
scenario, stars on the main-sequence which have been polluted by the
products of an earlier AGB generation have a higher helium
abundance. These polluted stars will then have a lower turnoff
mass for a given age than the unpolluted cluster stars with a
normal helium abundance, and thus will be more likely to end up on the
blue end of the HB. As in the deep-mixing scenario, the primordial
scenario also predicts a higher luminosity of the hydrogen shell on
the HB, due to the increased helium abundance.

Thus, both the deep mixing and primordial scenarios predict that the
stars with the strongest abundance anomalies should end up on the
bluest part of the HB, and have both a higher helium abundance and
higher luminosity (and lower gravity) than the HB stars with
normal abundances. 
Unfortunately, a straightforward test of these predictions is
complicated by the processes of diffusion and radiative levitation in
HB stars. The observed photospheric helium abundance in hot HB stars,
for example, has long been known to be strongly depleted (see Moehler
\cite{moeh01} for an overview) presumably due to gravitational settling.
Moreover, the use of luminosity or gravity discriminants for testing
these scenarios is complicated by the supersolar
iron abundances in HB stars hotter than $\approx$11,500~K 
(Glaspey et al. \cite{glmi89}, Behr et
al. \cite{beco99}, \cite{beco00}) due to radiative levitation (Michaud et
al. \cite{miva83}). If the hot HB stars with enhanced abundances are
analysed with model atmospheres at the cluster abundance, the stars
appear to have anomalously low gravities (Moehler et al.
\cite{mosw00}) or to be anomalously bright in certain bandpasses, such
as Str\"omgren $u$ (Grundahl et al. \cite{grca99}). However, when
Moehler et al. (\cite{mosw00}) used model atmospheres with the
correct iron abundance for the gravity determinations of hot HB stars
in NGC 6752, they found that the ``low-gravity" anomaly mostly
disappeared, although a small discrepancy remained for stars with
15,000 K $<$ \teff\ $<$ 20,000 K. Interestingly, Parise et
al. (\cite{pabo98}) also detect a possible luminosity offset of M13 HB
stars in this temperature range from their 1620 \AA\ photometry
obtained with the Ultraviolet Imaging Telescope (UIT).

Although selected hot HB stars in M13 have been observed for an
abundance analysis (Behr et al. \cite{beco99}) and studied in
Str\"omgren photometry (Grundahl et al. \cite{grca99}), there have
been no previous spectroscopic studies to obtain temperatures and
gravities. Here we present spectroscopy of 22 hot HB candidates in M13
to derive effective temperatures and surface gravities in order to
search for any deviations from canonical HB models.  We also estimate
abundances of helium, magnesium, and iron. Observations of four hot HB
stars in M 3 serve as a control sample.

\begin{figure*}[!ht]
\vspace{11.7cm}
%\special{psfile=m3m13bv.ps hscale=60 vscale=60 hoffset=0 voffset=350 
\includegraphics{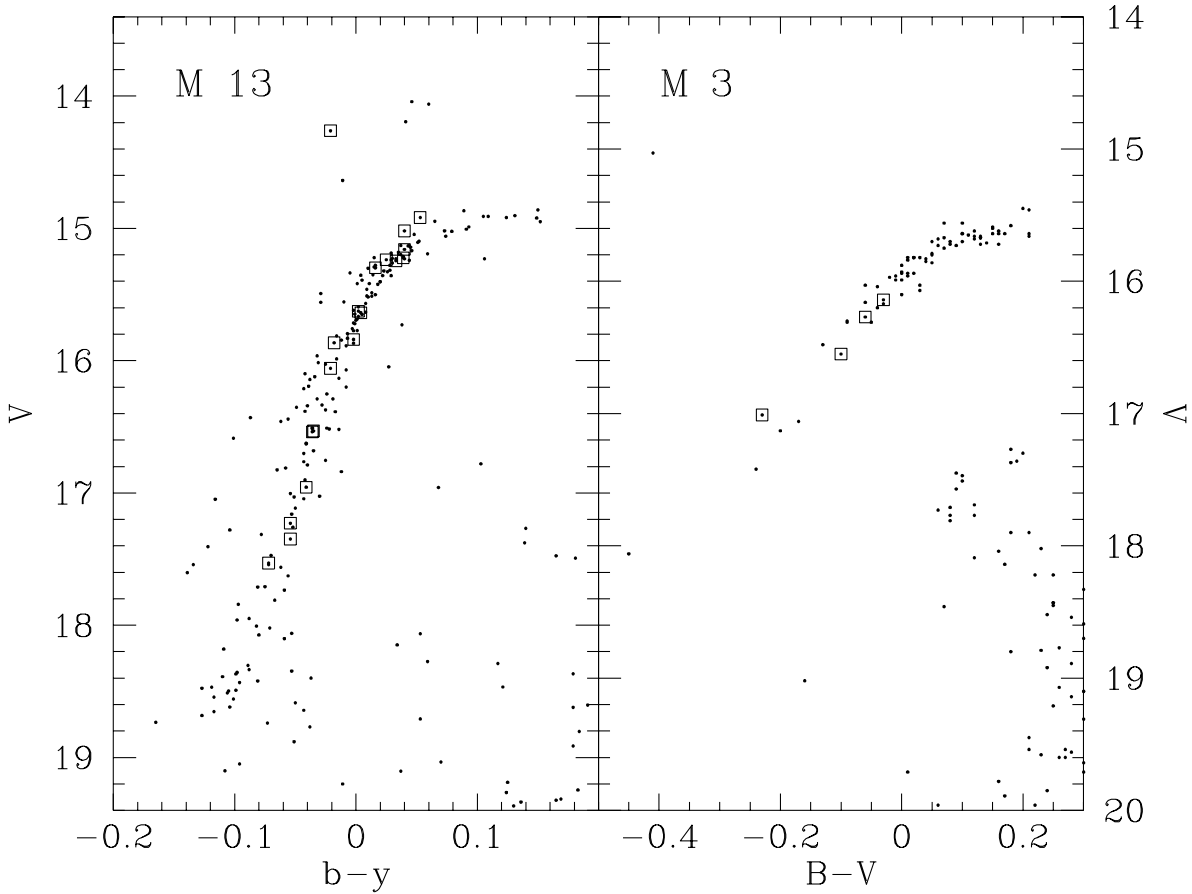}
\caption[]{The horizontal branches of M~13 (Grundahl et al. 1998, {\bf
left panel}) and M~3 (Buonanno et al. 1994, {\bf right panel}) with
the spectroscopic targets marked by open squares.
\label{m3_m13_cmd}}
\end{figure*}

\section{Observations\label{sec_obs}} 

We selected our targets in M~13 from
the Str\"omgren photometry of Grundahl et al. (\cite{grva98}, 
see Table~\ref{targets2}) and those in 
M~3 from the Johnson photometry of Buonanno et al. (\cite{buco94},
see Table~\ref{targets1}). We also included three targets in M~13 from
Behr et al. (\cite{beco99}) for comparison. Our targets are
plotted along the HBs of M~13 and M~3 in Fig. 1. For our
observations we used the Calar Alto 3.5m telescope with the TWIN
spectrograph. While we observed in both channels (blue and red) the 
crowding and consequent stray light of the primarily red cluster stars made 
the data from the red channel very difficult to reduce and analyse. We will 
therefore limit ourselves to the discussion of the data from the blue 
channel. There we used CCD\#11 (SITe\#12a,
800$\times$2000 pixels, (15~$\mu$m)$^2$ pixel
size, read-out noise 6~e$^-$, conversion factor 1.1~e$^-$/count) and
grating T12 (72~\AA mm$^{-1}$) to cover a wavelength range
of 3410~\AA\ -- 5570~\AA .
Combined with a slit width of 1.5\arcsec\ we thus achieved a mean spectral
resolution of 3.4~\AA\ as determined from the FWHM of the wavelength 
calibration lines. The spectra were obtained on June 4--6, 1999. For
calibration purposes we observed each night ten bias frames and ten dome
flat-fields with a mean exposure level of about 10,000~counts 
each. Before and
after each science observation we took HeAr spectra for wavelength
calibration purposes. We observed dark frames of 3600 and 1800 sec duration
to measure the dark current of the CCD. As flux standard stars we used
Feige~56, HZ~44, BD$+$28$^\circ$4211, and HZ~21.
\begin{table*}
\caption[]{Coordinates, photometric data, and heliocentric radial 
velocities for target stars in M~13
(Grundahl et al. 1998 [G]; Piotto et al.\ 2002 [WF3035, WF3485];
this paper [WF3085, WFPC2 data reduced as described by Dolphin 2000]). 
H$\beta$ indices with a large ($>$\magpt{0}{05}) uncertainty are marked 
with a colon.
\label{targets2}}
\begin{tabular}{llllllll}
\hline
\hline
Star & $\alpha_{2000}$ & $\delta_{2000}$ & $y$ & $b-y$ & $u-b$ & $\beta$ &
$v_{\rm rad}$\\
 & & & & & & & [km s$^{-1}$] \\
\hline
G43 & \RA{16}{41}{34}{7} & \DEC{+36}{29}{13}{7} & \magpt{14}{262} & 
\magpt{-0}{021} & \magpt{+0}{688} & \magpt{2}{651} & $-$287\\
G129 & \RA{16}{41}{41}{8} & \DEC{+36}{29}{20}{5} & \magpt{14}{918} & 
\magpt{+0}{053} & \magpt{+1}{632} & \magpt{2}{825} & $-$257\\
G137 & \RA{16}{41}{46}{3} & \DEC{+36}{30}{12}{6} & \magpt{15}{019} & 
\magpt{+0}{040} & \magpt{+1}{584} & \magpt{2}{837} & $-$251\\
G195 & \RA{16}{41}{43}{1} & \DEC{+36}{30}{08}{1} & \magpt{15}{159} & 
\magpt{+0}{040} & \magpt{+1}{554} & \magpt{2}{836} & $-$229\\
G201 & \RA{16}{41}{47}{9} & \DEC{+36}{29}{47}{3} & \magpt{15}{237} & 
\magpt{+0}{025} & \magpt{+1}{485} & \magpt{2}{832} & $-$276\\
G208 & \RA{16}{41}{34}{1} & \DEC{+36}{29}{51}{4} & \magpt{15}{298} & 
\magpt{+0}{016} & \magpt{+1}{347} & \magpt{2}{820}: & $-$244\\
G219 & \RA{16}{41}{39}{1} & \DEC{+36}{29}{54}{0} & \magpt{15}{222} & 
\magpt{+0}{039} & \magpt{+1}{564} & \magpt{2}{844} & $-$263\\
G235 & \RA{16}{41}{40}{1} & \DEC{+36}{29}{00}{1} & \magpt{15}{245} & 
\magpt{+0}{033} & \magpt{+1}{528} & \magpt{2}{843} & $-$245\\
G297 & \RA{16}{41}{35}{9} & \DEC{+36}{29}{27}{0} & \magpt{15}{626} & 
\magpt{+0}{002} & \magpt{+1}{210} & \magpt{2}{795} & $-$245\\
G314 & \RA{16}{41}{39}{9} & \DEC{+36}{30}{00}{5} & \magpt{15}{638} & 
\magpt{+0}{004} & \magpt{+1}{207} & \magpt{2}{804} & $-$268\\
G322 & \RA{16}{41}{40}{7} & \DEC{+36}{28}{52}{5} & \magpt{15}{865} & 
\magpt{-0}{018} & \magpt{+0}{711} & \magpt{2}{696} & $-$246\\
G342 & \RA{16}{41}{38}{8} & \DEC{+36}{30}{42}{4} & \magpt{15}{840} & 
\magpt{-0}{002} & \magpt{+1}{036} & \magpt{2}{766} & $-$231\\
G382 & \RA{16}{41}{41}{1} & \DEC{+36}{29}{39}{1} & \magpt{16}{058} & 
\magpt{-0}{021} & \magpt{+0}{649} & \magpt{2}{674} & $-$255\\
G496 & \RA{16}{41}{41}{1} & \DEC{+36}{30}{05}{7} & \magpt{16}{538} & 
\magpt{-0}{036} & \magpt{+0}{384} & \magpt{2}{673} & $-$231\\
G503 & \RA{16}{41}{33}{8} & \DEC{+36}{30}{15}{5} & \magpt{16}{531} & 
\magpt{-0}{035} & \magpt{+0}{532} & \magpt{2}{713}: & $-$221\\
G635 & \RA{16}{41}{40}{5} & \DEC{+36}{31}{19}{7} & \magpt{16}{957} & 
\magpt{-0}{041} & \magpt{+0}{316} & \magpt{2}{696}: & $-$246\\
G744 & \RA{16}{41}{48}{5} & \DEC{+36}{29}{59}{4} & \magpt{17}{229} & 
\magpt{-0}{054} & \magpt{+0}{266} & \magpt{2}{679}: & $-$251\\
G781 & \RA{16}{41}{48}{8} & \DEC{+36}{29}{28}{2} & \magpt{17}{348} & 
\magpt{-0}{054} & \magpt{+0}{240} & \magpt{2}{661}: & $-$237\\
G827 & \RA{16}{41}{41}{5} & \DEC{+36}{30}{37}{0} & \magpt{17}{530} & 
\magpt{-0}{072} & \magpt{+0}{168} & \magpt{2}{671} & $-$252\\
\hline
Star & $\alpha_{2000}$ & $\delta_{2000}$ & $V$ & $B-V$ & & & $v_{\rm rad}$\\
 & & & & & & & [km s$^{-1}$]\\
\hline
WF3035 & \RA{16}{41}{48}{1} & \DEC{+36}{28}{20}{6} & \magpt{15}{34} & 
\magpt{-0}{01} & & & $-$288\\
WF3085 & \RA{16}{41}{37}{8} & \DEC{+36}{26}{37}{1} & \magpt{16}{15} & 
\magpt{-0}{10} & & & $-$249\\
WF3485 & \RA{16}{41}{36}{9} & \DEC{+36}{26}{47}{7} & \magpt{16}{25} & 
\magpt{-0}{06} & & & $-$243\\
\hline
\end{tabular}
\end{table*}

\begin{table*}
\caption[]{Coordinates, photometric data, and heliocentric radial 
velocities for the target stars in M~3 
(Buonanno et al. 1994)\label{targets1}}
\begin{tabular}{llllll}
\hline
\hline
Star & $\alpha_{2000}$ & $\delta_{2000}$ & $V$ & $B-V$ & $v_{\rm rad}$\\
 & & & & & [km s$^{-1}$]\\
\hline
B254 & \RA{13}{42}{05}{3} & \DEC{+28}{25}{09} & \magpt{16}{14} & 
\magpt{-0}{03} & $-$151\\
B352 & \RA{13}{42}{28}{8} & \DEC{+28}{24}{33} & \magpt{17}{01} &
\magpt{-0}{23} & $-$149\\ 
B445 & \RA{13}{42}{33}{4} & \DEC{+28}{23}{57} & \magpt{16}{27} &
\magpt{-0}{06} & $-$146\\ 
B621 & \RA{13}{41}{56}{0} & \DEC{+28}{22}{54} & \magpt{16}{55} &
\magpt{-0}{10} & $-$177\\ 
\hline
\end{tabular}
\end{table*}

In order to observe as many stars as possible we oriented the slit to cover 
up to four hot HB stars at once. This of course did not allow us to reduce the 
light loss due to atmospheric dispersion by observing along the parallactic 
angle and also required observations in fairly crowded regions.

\section{Data Reduction\label{sec_redu}}

\begin{figure*}[h]
\vspace{22cm}
%\special{psfile=m3_m13_spec.ps hscale=80 vscale=80 hoffset=-13 voffset=-20 
\includegraphics{3155f2.ps2}
\caption[]{Normalized spectra of the programme stars in M~13 and M~3.
The part shortward of 3900~\AA\ was normalized by 
taking the highest flux point as continuum value. 
%The \ion{He}{i} lines $\lambda\lambda$ 4026~\AA, 4388~\AA, 4471~\AA,
%4922~\AA, and the \ion{Mg}{ii} line 4481~\AA\ are marked (if visible in the
%spectrum). 
\label{m3_m13_spec}}
\end{figure*}

We first averaged the bias and flat field frames separately for each
night. The mean bias frames of the second and the third night showed
the same level of about 923 counts (and were therefore averaged),
whereas the one from the first night had on average 877 counts. The
same difference was found in the overscan regions of the respective
bias frames. We therefore determined the mean overscan of each science
frame before deciding which bias frame to use (which was then adjusted
to the individual overscan level of the science frame). The mean dark
current determined from long dark frames showed no structure and
turned out to be negligible (1.5$\pm$2 counts/hr/pixel).

We determined the spectral energy distribution of the flat field lamp by
averaging the mean flat fields of each night along the spatial axis. These
one-dimensional ``flat field spectra" were then heavily smoothed and used
afterwards to normalize the dome flats along the dispersion axis. As the
normalized flat fields of each night differed slightly from each other we 
always corrected the science spectra using the flat field obtained for the 
same night.

For the wavelength calibration we fitted 2$^{\rm nd}$-order polynomials to
the dispersion relations of the HeAr spectra (using 46 unblended lines)
which resulted in mean
residuals of $\le$0.15\AA. We rebinned the frames two-dimensionally to
constant wavelength steps. Before the sky fit the frames were smoothed
along the spatial axis to erase cosmic ray hits in the background. To
determine the sky background we had to find regions without any stellar
spectra, which were sometimes not close to the place of the object's
spectrum. Nevertheless the flat field correction and wavelength calibration
turned out to be good enough that a linear fit to the spatial distribution
of the sky light allowed the sky background at the object's position to be
reproduced with sufficient accuracy. This means in our case that after the
fitted sky background was subtracted from the unsmoothed frame we do not
see any absorption lines caused by the predominantly red stars of the
clusters. The sky-subtracted spectra were extracted using Horne's
(\cite{horn86}) algorithm as implemented in MIDAS (Munich Image Data
Analysis System). 

Finally the spectra were corrected for atmospheric extinction using the
extinction coefficients for La Silla (T\"ug \cite{tueg77}) as 
implemented in MIDAS.
The data for the flux standard stars were
taken from Hamuy et al.\ (\cite{hamu92}, Feige~56) and Oke (\cite{oke90},
all others)
and the response curves were fitted
by splines or high-order polynomials. 
The flux-calibration is helpful for the later 
normalization of the spectra as it takes out all large-scale sensitivity 
variations of the instrumental setup. Atmospheric dispersion will cause 
light loss especially at blue end of the spectral range so that the flux 
distribution of the calibrated spectra cannot be used to infer temperatures 
or gravities (e.g. from the Balmer jump).

We derived radial velocities 
from the positions of the Balmer and helium lines. The
resulting heliocentric velocities are listed in Tables~\ref{targets2}
and \ref{targets1}. The error of the velocities (as estimated from
the scatter of the velocities derived from individual lines) is about
30~km~s$^{-1}$. The average radial velocity of the M~13 stars of
$-$251~km s$^{-1}$ agrees well with the literature value of $-$246~km
s$^{-1}$ (Lupton et al. \cite{lugu87}) within the error bars. The
same is true for M~3 with $-$156~km s$^{-1}$ (our result)
vs. $-$147~km s$^{-1}$ (Pryor et al. \cite{prla88}). Individual
velocities may deviate considerably from the mean value because the
stars may not have been in the center of the slit along the dispersion
axis during the observation due to the placement of the slit to
include several target stars simultaneously.

The Doppler-corrected spectra were then co-added and 
normalized by eye and are plotted in 
Fig.~\ref{m3_m13_spec}.

\section{Atmospheric Parameters\label{sec_par}}
\subsection{Photometric estimates}
To get a first estimate of effective temperatures and surface
gravities we used the Str\"omgren $uvby\beta$ photometry given in
Table~\ref{targets2}.

Because an empirical calibration of the Str\"omgren indices is not yet
available for hot HB stars, we used the Str\"omgren indices (including
H$\beta$) computed with the ATLAS9 code by Castelli (\cite{cast98}) to
derive values for \teff\ and log g. We adopted the model with [Fe/H] =
$-$1.5 with $\alpha$ elements enhanced by $+$0.4 dex, a solar helium
abundance and a microturbulent velocity of 1 km s$^{-1}$. The
normalization of the H$\beta$ indices by Castelli uses the values for
the Sun and Vega as anchor points for a linear interpolation relating
the calculated indices, $\beta_{calc}$ to the observed indices,
$\beta_{obs}$
$$
\beta_{obs} = 1.287\beta_{calc} + 2.5298
$$
The observed Str\"omgren $uvby$ photometry was dereddened by \ebv\ =
\magpt{0}{02}.   

For the stars with $b-y <$\magpt{0}{01} ($\stackrel{\wedge}{\approx}$11,000~K), 
the best-fit values of \teff\ and \logg\ were 
determined by minimizing the $\chi^2$ difference between the observed and 
calculated
$uvby$ and H$\beta$ photometry. For the stars with $b-y >$\magpt{0}{01}, we
computed theoretical ($a,r$) indices 
$$
a =  1.36 \cdot (b-y)+0.36\cdot m0  + 0.18\cdot c0 - 0.2448
$$
$$
r =  -0.07 \cdot (b-y)+0.35\cdot c0-(\beta-2.565)
$$
where for stars with 8500~K $<$ \teff $<$ 11,000~K $a$ correlates with \teff\
and $r$ correlates with \logg\ (Napiwotzki et al.\ \cite{nasc93}).
The ($a,r$) method was also used for field HB stars with similar
Str\"omgren colours by Kinman et al. (\cite{kica00}).  The values of
\teff\ and \logg\ derived from Str\"omgren photometry are given in
Table~\ref{m3_m13_par_m15a} (Cols. 2 and 3) and plotted in the upper
panel of Fig.~\ref{m3_m13_tg0}.  These values were then used as
starting points for the fits of model spectra to the observed Balmer
lines in order to derive effective temperatures and surface gravities
(see below). A prior idea of the probable range in temperature and
gravity is especially helpful for stars near the Balmer maximum, where
the fit of the spectral lines may yield different results depending on
the starting values, although the solutions on the two sides of the
Balmer maximum usually show quite different values of $\chi^2$. For
reasons of consistency we used the photometrically determined
parameters also as starting values for the hotter stars, although the
results of the line profile fitting depend very little on the starting
values for temperatures above about 12,000~K.

For those stars that do not have Str\"omgren photometry we used
$B-V$ from Tables~\ref{targets2} and \ref{targets1} to estimate
temperatures. As we only have $B-V$ we need additional information to
decide on a combination of effective temperature and surface
gravity. We therefore forced the parameters obtained from $B-V$ to
agree with the canonical zero-age HB model from Sweigart
(\cite{swei97b}) for [M/H] = $-$1.5. 
These parameters (listed in Table~\ref{m3_m13_par_m15a})
should therefore be taken with many
grains of salt and are useful mainly as starting points for the line
profile fitting. They are therefore not plotted in the top panel
of Fig.~\ref{m3_m13_tg0}. The $B-V$ colour of B~352 in M~3 is very probably
wrong, as it suggests an effective temperature of well above 25000~K,
in which case one would expect an absolute brightness of about
\magpt{+4}{2} instead of \magpt{+1}{9}.
\subsection{Line profile fitting}
We first computed model atmospheres using ATLAS9 (Kurucz
\cite{kuru93}) for [Fe/H] = $-$1.5 and [$\alpha$/Fe] = $+$0.4, using a
mixing length parameter of 1.25 and no overshoot.  This way our models
should be identical to those of Castelli (\cite{cast98}) used for the
analysis of the Str\"omgren photometry and also to those used by
Kinman et al. (\cite{kica00}) for the analysis of field horizontal
branch stars.
 Then we used Lemke's
version\footnote{For a description see
http://a400.sternwarte.uni-erlangen.de/$\sim$ai26/linfit/linfor.html}
of the LINFOR program (developed originally by Holweger, Steffen, and
Steenbock at Kiel University) to compute a grid of theoretical spectra
which include the Balmer lines H$_\alpha$ to H$_{22}$ and \ion{He}{i}
lines, but no metal lines. The grid covered the range
7,000~K~$\leq$~\teff~$\leq$~15,000~K, 2.5~$\leq$~\logg~$\leq$~5.5 at a
solar helium abundance \loghe=$-1.0$.
We also computed a grid with identical abundances, but no
convection, for 7,000~K~$\leq$~\teff~$\leq$~35,000~K, 
2.5~$\leq$~\logg~$\leq$~5.5. This grid was used for all stars
with photometric temperatures above 12,000~K, where convection should
have vanished.

\begin{figure}
\vspace{11.5cm}
%\special{psfile=m3_m13_tg0k.ps hscale=42 vscale=42 hoffset=-0 voffset=-5
\includegraphics{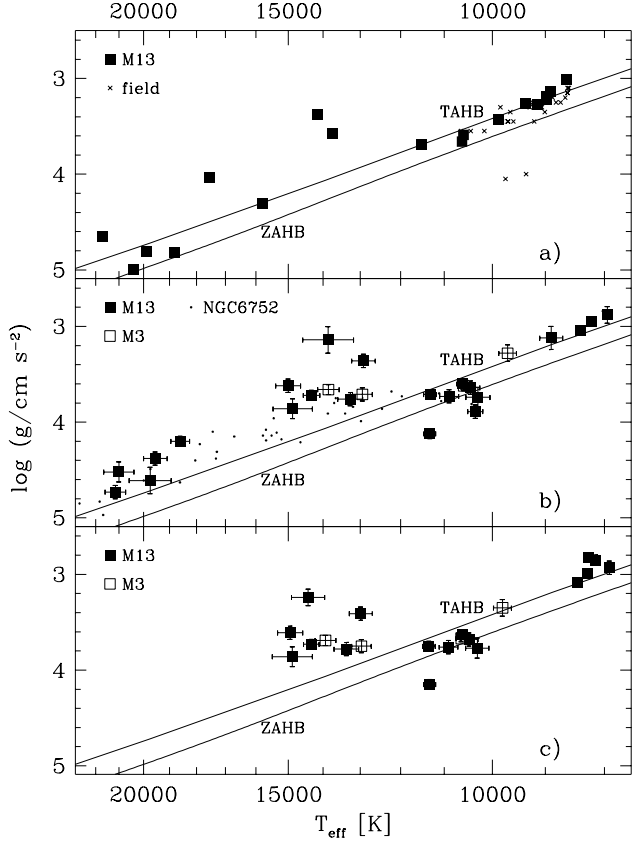}
\caption[]{
Atmospheric parameters of the programme stars in M~3 and M~13
obtained with $\alpha$-enhanced metal-poor model atmospheres
([Fe/H] = $-$1.5, [$\alpha$/Fe] $+$0.4).
The solid lines mark the zero-age (ZAHB) and terminal-age
(TAHB) loci of canonical HB tracks for [M/H] = $-$1.54. These loci
define the region within which the HB models spend 99\% of their HB
lifetime.   
\\{\bf a)} Temperatures and gravities from Str\"omgren
$uvby\beta$ photometry (for comparison we also show results for
cool field BHB stars obtained with the same method by Kinman et
al. \cite{kica00}). 
\\ {\bf b)} Temperatures and gravities
for all stars are derived from fits to the Balmer line profiles, using
as starting values the photometrically determined parameters. The
error bars note the formal 1 $\sigma$ errors from the line profile
fits. The model spectra used here do not include metal lines.
For comparison we also show the results for blue tail stars in NGC~6752
from Moehler et al. (\cite{mosw00}, obtained with metal-poor model spectra 
without metal lines).
\\ {\bf c)}
Temperatures and gravities of the programme stars cooler than 15,000~K
from Balmer line
profile fits using model spectra that include metal lines for [Fe/H] =
$-$1.5 and [$\alpha$/Fe] = $+$0.4.
\label{m3_m13_tg0}} 
\end{figure}

In Table~\ref{m3_m13_par_m15a} (Cols. 5 and 6) we list the
results obtained from fitting the Balmer lines H$_\beta$ to H$_{10}$
(excluding H$_\epsilon$ to avoid the \ion{Ca}{ii}~H line) with these
model spectra. To establish the best fit, we used the routines
developed by Bergeron et al.\ (\cite{besa92}) and Saffer et al.\
(\cite{sabe94}), as modified by Napiwotzki et al. (\cite{nagr99}),
which employ a $\chi^2$ test. The $\sigma$ necessary for the
calculation of $\chi^2$ is estimated from the noise in the continuum
regions of the spectra. The fit program normalizes model spectra {\em
and} observed spectra using the same points for the continuum
definition. The errors given in Tables~\ref{m3_m13_par_m15a} and
\ref{m3_m13_par_p05} are r.m.s. errors derived from the $\chi^2$ fit
(see Moehler et al.\ \cite{mosw99} for more details). These errors are
obtained under the assumption that the only error source is
statistical noise (derived from the continuum of the
spectrum). However, errors in the normalization of the spectrum,
imperfections of flat field/sky background correction, variations in
the resolution (e.g. due to seeing variations when using a rather
large slit width) and other effects may produce systematic rather than
statistical errors, which are not well represented by the error
obtained from the fit routine. 
%
%In order to avoid any biases due to
%normalization we used the flux calibrated spectra for the fit and not
%the normalized ones. This should ensure that the observed lines are
%normalized in exactly the same way as the model spectra. Tests with
%the normalized spectra plotted in Fig.~\ref{m3_m13_spec} show,
%however, that the differences are at most 2\% in \teff\ (on average
%below 0.1\%) and 0.01 dex in \logg\ (after correcting for the effect
%of the different temperature, on average no difference in \logg\ was
%detected).  
%
As the helium abundance is fixed in the model spectra we
did not try to fit any helium lines, because this would introduce
strong biases in case of non-solar helium abundances (expected for the
hot HB stars, see Sect.~\ref{ssec_iron}). The results of this analysis
are plotted in Fig.~\ref{m3_m13_tg0} (central panel). As it is
not precisely clear at which temperature convection truly vanishes in
HB stars we fitted stars with photometric temperatures between 9,000~K
and 12,000~K with both the convective and the non-convective model
spectra. The non-convective model spectra yield effective temperatures
lower by on average 10~K and surface gravities lower by on average
0.015 dex. Thus the effect of convection at these temperatures is
certainly negligible for our analysis.

\begin{table*}
\caption[]{Atmospheric parameters for our programme stars in M~3 (B)
and M~13 (G, WF) as derived from photometry (Cols. 2 and 3) and fits to
the Balmer line profiles using metal-poor, $\alpha$-enhanced model
atmospheres with solar helium abundances.  Cols. 5 and 6 list the
results obtained with model spectra which contain only hydrogen and
helium lines, while Cols. 7 and 8 give the results obtained with
model spectra that also include metal lines for [Fe/H] = $-$1.5 and
[$\alpha$/Fe] = $+$0.4.
For those stars where we could obtain a
meaningful fit we also obtained iron abundances from spectrum
synthesis (Cols. 4 and 9, see text for details). Given are abundances using
microturbulent velocities of 3 km s$^{-1}$ and (in brackets) for 0 km
s$^{-1}$.  The errors are statistical errors only.
\label{m3_m13_par_m15a}}
\begin{tabular}{lrrrrrrrr}
\hline
\hline
Star & \multicolumn{2}{c}{Photometry} &
log $\epsilon_{\rm Fe}$ & \multicolumn{4}{c}{Line profiles}
& log $\epsilon_{\rm Fe}$ \\
 & \teff & \logg & & \teff & \logg & \teff & \logg & \\
     & [K] & [cm s$^{-2}$] & & [K] & [cm s$^{-2}$] & [K] & [cm s$^{-2}$] & \\
\hline
G43  & 13050  & 2.60 & & 13900$\pm$710 & 3.14$\pm$0.14 
& 14400$\pm$450 & 3.24$\pm$0.09 & --\\
G129   & 8630 & 3.01 & 5.7 (5.6) & 8000$\pm$\hspace*{4.7pt}70 & 2.88$\pm$0.09
& 7900$\pm$\hspace*{4.7pt}50 & 2.93$\pm$0.07 & 5.1 (5.1)\\
G137   & 8910 & 3.14 & 6.4 (6.8) & 8400$\pm$\hspace*{4.7pt}50  & 3.04$\pm$0.02
& 8400$\pm$\hspace*{4.7pt}50 & 3.08$\pm$0.02 & 6.0 (6.4) \\
G195   & 8980 & 3.19 & 5.8 (5.8) & 8200$\pm$\hspace*{4.7pt}40  & 2.95$\pm$0.03
& 8300$\pm$\hspace*{4.7pt}40 & 2.99$\pm$0.02 & 5.2 (5.2) \\
G201   & 9360 & 3.26 & 6.4 (6.6) & 8900$\pm$200 & 3.12$\pm$0.12 &
8300$\pm$\hspace*{4.7pt}50 & 2.82$\pm$0.03 & 5.5 (5.6)\\
G208   & 9870 & 3.43 & 6.6 (6.6) & 10900$\pm$200 & 3.73$\pm$0.07 &
10900$\pm$200 & 3.76$\pm$0.07 & 6.9 (7.0)\\
G219   & 8970 & 3.22 & 5.8 (6.0) & 10300$\pm$160 & 3.89$\pm$0.07 &
8100$\pm$\hspace*{4.7pt}60 & 2.85$\pm$0.05 & 4.9 (5.2) \\
G235   & 9150 & 3.27 & 5.9 (5.8) & 11300$\pm$130 & 4.12$\pm$0.05 &
11300$\pm$130 & 4.15$\pm$0.05 & 6.8 (6.7)\\
G297  & 10590 & 3.59 & & 10600$\pm$130 & 3.60$\pm$0.05 &
10600$\pm$120 & 3.63$\pm$0.05 & 6.5 (6.6)\\
G314  & 10620 & 3.66 & & 10400$\pm$190 & 3.64$\pm$0.07 &
10500$\pm$180 & 3.68$\pm$0.07 & 6.0 (6.0)\\
G322  & 13750 & 3.58 & & 15000$\pm$370 & 3.62$\pm$0.07 &
14900$\pm$360 & 3.61$\pm$0.07 & 7.4 (7.5)\\
G342  & 11520 & 3.69 & & 11300$\pm$140 & 3.71$\pm$0.05 &
11300$\pm$140 & 3.75$\pm$0.05 & 6.2 (6.1)\\
G382  & 14160 & 3.38 & & 14300$\pm$240 & 3.72$\pm$0.05 &
14300$\pm$240 & 3.73$\pm$0.05 & 8.0 (8.0)\\
G496  & 17560 & 4.03 & & 18600$\pm$350 & 4.20$\pm$0.05 & & & --\\
G503  & 15800 & 4.31 & & 14900$\pm$580 & 3.86$\pm$0.10 &
14900$\pm$570 & 3.86$\pm$0.10 & 7.4 (7.6)\\
G635  & 18800 & 4.82 & & 19500$\pm$450 & 4.38$\pm$0.07 & & & --\\
G744  & 19880 & 4.81 & & 21000$\pm$620 & 4.52$\pm$0.10 & & & --\\
G781  & 20390 & 5.00 & & 19700$\pm$820 & 4.61$\pm$0.14 & & & --\\
G827  & 21720 & 4.65 & & 21200$\pm$430 & 4.73$\pm$0.07 & & & --\\
WF3035  & 9550 & 3.51 & & 10300$\pm$240 & 3.74$\pm$0.10 &
10300$\pm$240 & 3.77$\pm$0.10 & --\\
WF3085 & 10260 & 3.66 & & 12900$\pm$300 & 3.36$\pm$0.07 &
12900$\pm$300 & 3.41$\pm$0.07 & 7.5 (8.2)\\
WF3485  & 9790 & 3.57 & & 13300$\pm$330 & 3.76$\pm$0.07 &
13400$\pm$330 & 3.78$\pm$0.07 & 7.9 (8.4)\\
B254   & 9630 & 3.53 & & 10600$\pm$110 & 3.62$\pm$0.05 &
10600$\pm$110 & 3.67$\pm$0.03 & 6.3 (6.2)\\
B352  & 28600 & 5.63 & & 13900$\pm$300 & 3.66$\pm$0.05 &
13900$\pm$300 & 3.69$\pm$0.05 & 8.1 (8.6)\\
B445   & 9850 & 3.58 & & 9700$\pm$170 & 3.28$\pm$0.09 &
9800$\pm$170 & 3.35$\pm$0.09 & 6.2 (6.2)\\
B621  & 10170 & 3.64 & & 12900$\pm$260 & 3.71$\pm$0.07 &
13000$\pm$250 & 3.75$\pm$0.07 & 7.6 (7.9)\\
\hline
\end{tabular}
\end{table*}

\subsubsection{Stars hotter than 12,000~K}
The star G43 is found to have a high temperature (\teff\ $\approx$ 14,000 K)
which, along with its bright visual magnitude ($y$ = \magpt{14}{26}),
suggests that it is a UV-bright star. The star is clearly seen as such
on the 1600 \AA\ image of M13 obtained with the Ultraviolet Imaging
Telescope (UIT, Parise et al. \cite{pabo98}), but the crowding prevents a good
determination of its ultraviolet magnitude. Using the $y$ magnitude
and the derived effective temperature, and assuming an apparent
distance modulus of $(m-M)_V$ = \magpt{14}{51} toward M~13 and a
reddening of \Ebv\ = \magpt{0}{02}, we estimate a luminosity of $\log$
L/\Lsolar = 2.47. The evolutionary status of G43 is uncertain, but
given the large extreme HB (EHB) population in M13, it may be either a
He-shell burning post-EHB star, or a post-early AGB star (cf. Moehler
et al.\ \cite{mola98}). In any case, since G43 is almost certainly not
an HB star, it is excluded from further discussion in this paper.

The results from the Balmer line profile analyses place the stars
hotter than 12,000~K further away from the ZAHB than the results from
the Str\"omgren colours (see Fig.~\ref{m3_m13_tg0}, central
panel). These stars, however, show also weaker \ion{He}{i} lines than
predicted by model atmospheres with solar helium abundance, while in
the cooler stars the He I lines - where detectable - are well
described by line profiles for solar helium abundance.  The offset
from the canonical tracks seen in Fig.~\ref{m3_m13_tg0} for the stars
hotter than about 12,000~K in both M~3 and M~13 is already well known
 from other globular clusters (see, e.g. Moehler \cite{moeh01}). To
better illustrate this general behavior, we also give the results for
the blue HB stars in NGC~6752 obtained with similar methods by Moehler
et al. (\cite{mosw00}). Note that the gravities of the blue HB stars
in M~3, M~13 and NGC~6752 all follow the same trend with
\teff. Possible reasons for this behaviour are discussed in
Sects.~\ref{ssec_iron} and \ref{sec-he}.

\subsubsection{Stars cooler than 12,000~K}
As can be seen from the top and central panel in Fig.~\ref{m3_m13_tg0}
and from Table~\ref{m3_m13_par_m15a}, there is an excellent agreement
between the values of \teff\ and \logg\ derived from Str\"omgren
photometry, and those derived from a detailed Balmer line fit for the
three stars (G297, G314, G342) in M13 with Str\"omgren temperatures
between 10,000~K and 12,000~K. This agreement provides evidence in
support of the Str\"omgren calibration of Castelli (\cite{cast98}) for
deriving the atmospheric parameters of low-metallicity hot HB stars in
this temperature range. The atmospheric parameters of these stars as
well as B254 also agree well with canonical zero-age HB (ZAHB).

As can be seen in the top panel of Fig.~\ref{m3_m13_tg0}, 
the stars with Str\"omgren
temperatures cooler than $\approx$10,000~K show a small offset
towards low gravities from the canonical HB.  Fig.~\ref{m3_m13_cmd}
shows that at least the two coolest stars might actually be somewhat
evolved as they are brighter than the main HB in that region of
colour-magnitude diagram. There is also good agreement with gravities
derived for field blue HB stars in this temperature range by Kinman et
al.  (\cite{kica00}).

Comparing the top and central panel of Fig.~\ref{m3_m13_tg0}, however,
shows that there are significant discrepancies between the results
obtained from photometric and from spectroscopic data for these stars
(cf. Table~\ref{m3_m13_par_m15a}).  The results from the Balmer line
profile fitting move the cool BHB stars to hotter and cooler
temperatures, thereby producing a gap between about 9000~K to
10,000~K.

While these stars are not the central target of our investigation we
followed the suggestions of the referee to look for causes that might
explain this behaviour. Readers interested in the hot stars in M~3 and
M~13 may skip the remainder of this section and move directly to
Sect.~\ref{ssec_iron}.

The Balmer line profile fitting routine could not find a a model
spectrum that provides a good fit to all Balmer lines in two cases
(G235 and WF3035, both below the ZAHB in Fig.~\ref{m3_m13_tg0}):
Models at about 10,000~K to 11,000~K predict too weak H$_\delta$ to
H$_\beta$ lines, whereas models at about 8000~K predict too weak
H$_{10}$ to H$_8$ lines. The spectra of all other cool stars are very
well reproduced by the model fits.
  We checked the following possible
causes for problems in our Balmer line fitting:
\begin{description}
\item {\it Synthetic Spectra:} We compared 
``our'' temperature-pressure distribution and resulting theoretical 
Balmer profiles for selected
models with those computed by Castelli (available on http://kurucz.harvard.edu/)
and those computed using the program SYNSPEC (Hubeny \& Lanz
\cite{hula03}) and found no discernible differences.
\item [{\it Metal lines:}] The Balmer line fitting procedure we used
was originally developed for hot white dwarfs, which show only
hydrogen and/or helium lines. Blue HB stars below 10,000~K, however,
may show strong metal lines even in case of metal-poor stars, which
may affect the shape of the Balmer lines especially when observed at
low resolution.  To rule out the influence of metal lines on the line
profile fitting for the cooler stars we produced a grid of model
spectra with [Fe/H] = $-$1.5 and [$\alpha$/Fe] = $+$0.4 (with and
without convection for the same temperature ranges as above), which include
metal lines (up to effective temperatures of 15,000~K) and which we
used for stars with photometric temperature estimates below
15,000~K. The results of fits with these model atmospheres are plotted
in the bottom panel of Fig.~\ref{m3_m13_tg0} and listed in
Table~\ref{m3_m13_par_m15a} (Cols. 7 and 8). Two stars show large
changes (G201 moves from 8900~K to 8300~K, G219 moves from 10300~K to
8100~K) and the $\chi^2$ of all stars is significantly
reduced. However, the discrepancies with the evolutionary tracks
persist as well as the bad fits for G~235 and WF~3035.
\item [{\it Signal-to-Noise and/or resolution:}] Our data have S/N between 40
and 60 (determined from the continuum between 4500~\AA\ and 4600~\AA)
and a resolution of 3.4\AA. We convolved a theoretical spectrum from
our grid with metal lines to this resolution and added noise to
achieve a S/N of 40 and 60, respectively. Fitting this spectrum
reproduced the original parameters (9000~K, \logg = 3.0) to within
$\pm$100K in \teff\ and $\pm$0.02 dex in \logg.

Assuming that the real resolution might be better than determined from
the wavelength calibration frames (which illuminate the whole slit of
1\farcs5) we assumed a resolution of 3.0\AA\ for another test, both
with the theoretical model spectrum as with our data. Again the effect
is small, with changes of less than 50~K in \teff\ and 0.02 dex in
\logg. We also verified that the shape of the wavelength calibration
lines is well described by a Gaussian.
\item [{\it Rotation:}] Assuming a projected rotational velocity of 20
km s$^{-1}$ does not affect the results, which is not unexpected
considering the limited resolution of our spectra.
\end{description} 

So we currently have no explanation for the disagreement between the
atmospheric parameters for the cool stars derived from Str\"omgren photometry,
and those derived from Balmer-line fitting.
\begin{figure}[h]
\vspace{11cm}
%\special{psfile=iron_spec.ps hscale=40 vscale=40 hoffset=0 voffset=0 
\includegraphics{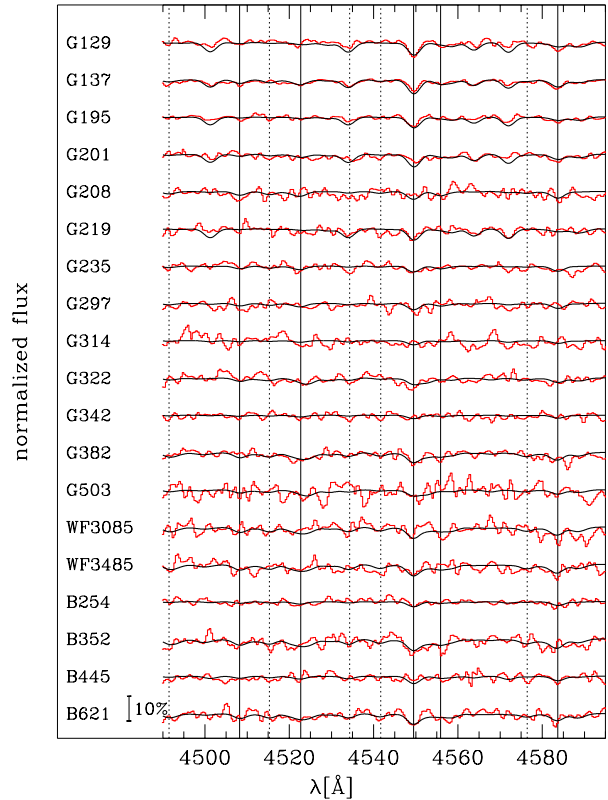}
\caption[]{Normalized spectra of the programme stars in M~13 and M~3 and
theoretical spectra for the atmospheric parameters (from the Balmer
line profiles) and iron abundances
listed in Table~\ref{m3_m13_par_m15a}. For stars hotter than 12,000~K we
show the fits obtained for atmospheric parameters from metal-rich
model atmosphere fits.
The solid and dotted lines mark the strongest and moderately strong
iron lines in the spectra. \label{m3_m13_ironspec}}
\end{figure}

\begin{table*}
\caption[]{Atmospheric parameters for our programme stars in M~3 (B)
and M~13 (G, WF) as derived from
fits to Balmer and helium lines using enriched scaled-solar model
spectra ([M/H] = $+$0.5). Cols. 2--4 give the results obtained with
model spectra containing only hydrogen and helium lines, Cols. 6--8
give the results obtained with model spectra including metal lines. 
For those stars where we could obtain a
meaningful fit we also obtained iron abundances from spectrum
synthesis (Col. 5, see text for details). Given are abundances using
microturbulent velocities of 3 km s$^{-1}$ and (in brackets) for 0 km
s$^{-1}$.  The errors are statistical errors only.
\label{m3_m13_par_p05}}
\begin{tabular}{lrrrrrrr}
\hline
\hline
Star & \multicolumn{3}{c}{Line profiles w/o metal lines} & log 
$\epsilon_{\rm Fe}$  & \multicolumn{3}{c}{Line profiles w metal lines} \\
 & \teff & \logg & \loghe & & \teff & \logg & \loghe \\
     & [K] & [cm s$^{-2}$] & & & [K] & [cm s$^{-2}$] & \\
\hline
%G43 & 13700$\pm$\hspace*{4.7pt}770 & 3.30$\pm$0.19 & $-$1.56$\pm$0.31 &
%& 13700$\pm$800  & 3.37$\pm$0.19  & $-$1.63$\pm$0.33\\
 G322 &
14600$\pm$\hspace*{4.7pt}400 & 3.82$\pm$0.09 & $-$2.32$\pm$0.23 & 7.4 (7.5)
& 14500$\pm$400  & 3.84$\pm$0.09  & $-$2.41$\pm$0.28\\
 G382 &
14100$\pm$\hspace*{4.7pt}200 & 3.94$\pm$0.05 & $-$2.07$\pm$0.12 & 7.8 (8.2) 
& 14100$\pm$300  & 4.00$\pm$0.07  & $-$2.13$\pm$0.14\\
 G496 &
17700$\pm$\hspace*{4.7pt}400 & 4.33$\pm$0.07 & $-$1.90$\pm$0.09 & -- 
& 17300$\pm$400 & 4.35$\pm$0.07 & $-$1.92$\pm$0.09\\
 G503 &
15200$\pm$\hspace*{4.7pt}600 & 4.20$\pm$0.12 & $-$2.60$\pm$0.42 & 7.4 (7.6)
& 15100$\pm$500  & 4.22$\pm$0.12  & $-$2.72$\pm$0.48\\
 G635 &
19200$\pm$\hspace*{4.7pt}600 & 4.58$\pm$0.09 & $-$2.15$\pm$0.10 & --
& 18700$\pm$600 & 4.59$\pm$0.09 & $-$2.16$\pm$0.12\\
 G744 &
20200$\pm$1100 & 4.61$\pm$0.12 & $-$2.04$\pm$0.12 & -- 
& 18400$\pm$700 & 4.57$\pm$0.12 & $-$1.94$\pm$0.12\\
 G781 &
18800$\pm$1000 & 4.71$\pm$0.16 & $-$1.46$\pm$0.16 & --
& 17800$\pm$800 & 4.70$\pm$0.16 & $-$1.37$\pm$0.17\\
 G827 &
20800$\pm$\hspace*{4.7pt}700 & 4.89$\pm$0.09 & $-$2.37$\pm$0.10 & --
& 19800$\pm$600 & 4.89$\pm$0.09 & $-$2.28$\pm$0.12\\
WF3085 & 
13100$\pm$\hspace*{4.7pt}300 & 3.65$\pm$0.09 & $-$2.34$\pm$0.35 & 7.4 (7.9)
& 12800$\pm$300  & 3.62$\pm$0.09  & $-$2.36$\pm$0.38\\
WF3485 & 
13300$\pm$\hspace*{4.7pt}300 & 4.05$\pm$0.09 & $-$2.31$\pm$0.38 & 7.7 (8.2)
& 13300$\pm$300  & 4.08$\pm$0.09  & $-$2.36$\pm$0.45\\
\hline
B352  &
13800$\pm$\hspace*{4.7pt}300 & 3.92$\pm$0.07 & $-$2.21$\pm$0.23 & 8.1 (8.5)
& 13800$\pm$300  & 3.99$\pm$0.07  & $-$2.29$\pm$0.29\\
B621  & 
13000$\pm$\hspace*{4.7pt}300 & 3.98$\pm$0.07 & $-$2.20$\pm$0.31 & 7.4 (7.7)
& 12800$\pm$300  & 3.98$\pm$0.09  & $-$2.34$\pm$0.36\\
\hline
\end{tabular}
\end{table*}

\begin{table}
\caption[]{Magnesium abundances for our
programme stars in M~3 (B) and M~13 (G, WF) for microturbulent
velocities of 3 km s$^{\rm -1}$ and (in brackets) 0 km s$^{-1}$. We
also give the atmospheric parameters used for the abundance determinations.
\label{m3_m13_mg}}
\begin{tabular}{lrrrr}
\hline
\hline
Star & \teff & \logg & \loghe & log $\epsilon_{\rm Mg}$ \\
     & [K] & [cm s$^{-2}$] & & 3 km/s (0 km/s)\\
\hline
%G129 &   7900 & 2.93 & $-$1.00 & 5.9 (6.1)\\
%G137 &   8400 & 3.08 & $-$1.00 & 6.1 (6.4)\\
%G195 &   8300 & 2.99 & $-$1.00 & 6.1 (6.5)\\
%G201 &   8900 & 3.12 & $-$1.00 & 6.3 (6.6)\\
%G219 &   8100 & 2.85 & $-$1.00 & 6.5 (6.9)\\
%G235 &  11300 & 4.15 & $-$1.00 & 6.3 (6.5)\\
G297 &  10600 & 3.63 & $-$1.00 & 6.2 (6.5)\\
G314 &  10400 & 3.68 & $-$1.00 & 6.0 (6.2)\\
G322 &  14600 & 3.82 & $-$2.32 & 6.2 (6.3)\\
G342 &  11300 & 3.75 & $-$1.00 & 5.9 (6.2)\\
G382 &  14100 & 3.94 & $-$2.07 & 6.1 (6.3)\\
G496 &  17700 & 4.33 & $-$1.90 & 6.2 (6.3)\\
G503 &  15200 & 4.20 & $-$2.60 & 6.7 (6.9) \\
G635 &  19200 & 4.58 & $-$2.15 & 6.9 (7.0) \\
WF3085 & 13100 & 3.65 & $-$2.34 & 6.3 (6.6)\\
WF3485 & 13300 & 4.05 & $-$2.31 & 6.0 (6.1)\\
\hline
B254  &  10600 & 3.67 & $-$1.00 & 6.4 (6.7)\\
B352  &  13800 & 3.92 & $-$2.21 & 6.1 (6.2)\\
B445  &   9800 & 3.35 & $-$1.00 & 6.0 (6.3)\\
B621  &  13000 & 3.98 & $-$2.20 & 6.2 (6.4)\\
\hline
\end{tabular}
\end{table}

\begin{figure}
\vspace{11.5cm} 
%\special{psfile=plot_abu.ps hscale=42 vscale=42 hoffset=-0 voffset=-5 angle=0}
\includegraphics{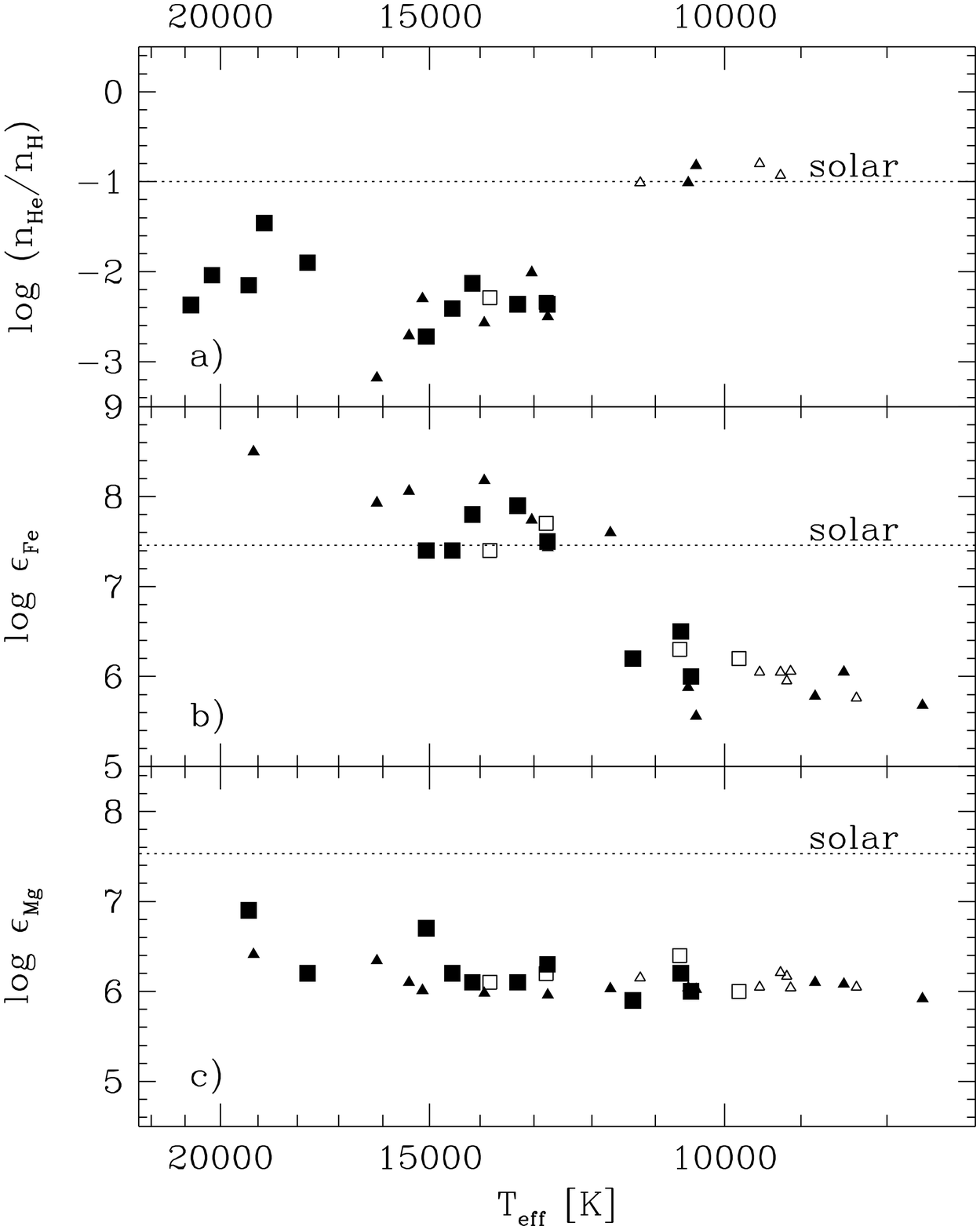}
\caption[]{Abundances of {\bf a)} helium, {\bf b)} iron, and {\bf c)}
magnesium for the programme stars in M~3 (open squares) and M~13 
(filled squares). 
Also given are the results of Behr (priv. comm.) for
stars in M~3 (open triangles) and M~13 (filled triangles). 
\label{m3_m13_abu}} 
\end{figure}

\section{Radiative levitation of heavy elements\label{ssec_iron}}

As described in Moehler et al.\ (\cite{mosw00}) and Behr et al.
(\cite{beco99}, \cite{beco00}), hot HB stars show helium depletion and
iron enrichment for effective temperatures above 11,000~K to
12,000~K. To verify this behaviour from our low-resolution spectra, we
performed a spectrum synthesis to reproduce the iron lines in the
wavelength region 4490 -- 4590 \AA\ (as described in Moehler et al.\
\cite{mosw00}). First estimates are obtained using the atmospheric
parameters given in Cols. 7 and 8 of Table~\ref{m3_m13_par_m15a} and
are listed in column 9 of the same table. The best fitting theoretical
spectra are shown in Fig.~\ref{m3_m13_ironspec} together with the
observed spectra.  While it is obvious from Fig.~\ref{m3_m13_ironspec}
that the cool stars show significant iron lines our fits (using the
atmospheric parameters from the Balmer line profile fits) indicate
very low iron abundances. If we use the atmospheric parameters derived
from Str\"omgren photometry instead the iron abundances are much
closer to [Fe/H] $\approx -1.5$ (see column 4 of
Table~\ref{m3_m13_par_m15a}). We take this as further evidence that
the results from the Balmer line profile fits cannot be trusted for
stars cooler than about 9500~K and we will omit these stars (including
WF~3035 due to its fit problems) from all further discussion, as they
are not the central goal of this paper.

   In order to be at least roughly consistent with the actual stellar
abundances, we redetermined \teff, \logg\ and \loghe\ for the stars
showing evidence for enriched iron abundances using model atmospheres
with super-solar metallicity ([M/H] = $+$0.5), varying helium
abundance and no convection. 
This was done by fitting the Balmer lines H$_\beta$ to
H$_{10}$ (excluding H$_\epsilon$ to avoid the \ion{Ca}{ii} H line) and
the \ion{He}{i} lines 4026 \AA, 4388 \AA, 4471 \AA, and 4921 \AA\ with
these metal-rich model atmospheres. The results are given in
Cols. 2--4 of Table~\ref{m3_m13_par_p05}. Using these new atmospheric
parameters, we then repeated the iron abundance analysis and used
those fits for Fig.~\ref{m3_m13_abu}.

We also determined magnesium abundances where possible from the
equivalent width of the \ion{Mg}{ii} line at 4482\AA\ (see
Table~\ref{m3_m13_mg}).

\begin{figure}[h]
\vspace{7.3cm}
%\special{psfile=m13uuy.ps hscale=40 vscale=40 hoffset=0 voffset=-65 
\includegraphics{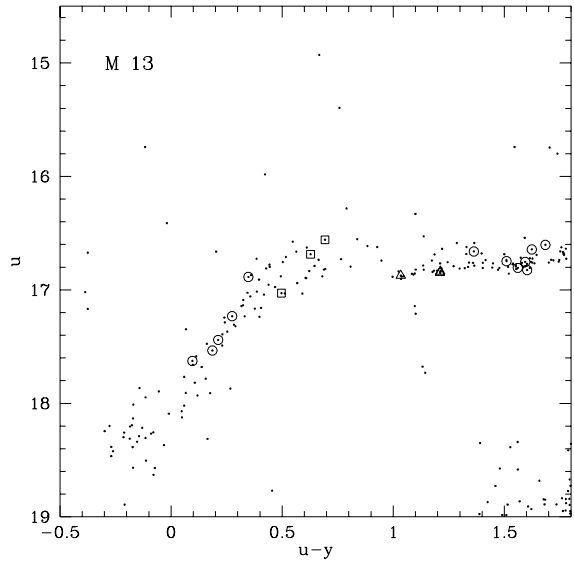}
\caption[]{The $u$-jump in M~13 with the spectroscopic targets marked:
Open triangles mark stars with low iron abundance ([Fe/H] $\approx
-1.5$), open squares mark stars with high iron abundance ([Fe/H]
$\approx +0.5$). Stars for which the iron abundance could not be
determined due to the limited S/N (hot stars) or uncertainties in the
atmospheric parameters (cool stars) are marked by open circles.
\label{m13uy}}
\end{figure}

   The final iron abundances derived from our spectra are listed in
Col. 9 of Table~\ref{m3_m13_par_m15a} for stars cooler than 12,000 K
and in Col. 5 of Table~\ref{m3_m13_par_p05} for stars hotter than
12,000 K and are plotted in the central panel of
Fig.~\ref{m3_m13_abu}. For consistency we always plot the abundances
obtained with a microturbulent velocity of 3 km s$^{-1}$, although 0
km s$^{-1}$ would be more appropriate for the stars affected by
diffusion.  A drastic change in the iron abundance between 11,500 K
and 13,000 K is obvious, in good agreement with the findings of
Glaspey et al. (\cite{glmi89}) for two hot HB stars in NGC 6752 and
Behr et al. (\cite{beco99}, \cite{beco00}) for hot HB stars in M 13
and M 15. The results in Fig.~\ref{m3_m13_abu} show that all stars
hotter than 12,000 K for which we can estimate the iron abundance show
evidence for radiative levitation. The iron abundance for the hotter
stars is a factor of 50--100 greater than that of the cluster in
general and consistent with that required to explain the
Stromgren $u$-jump discussed by Grundahl et al. (\cite{grca99}, $\log
\epsilon_{\rm Fe}$ = 8.1).  We also find a one-to-one correspondence
between the position of a star relative to the $u$-jump
(blueward/redward) and its iron abundance (enhanced/cluster abundance)
whenever we were able to determine an iron abundance
(cf. Fig.~\ref{m13uy}).

The upper panel of Fig.~\ref{m3_m13_abu} shows that the onset of
radiative levitation is accompanied by a drop in the helium abundance
by 1 dex or more. As the \ion{He}{i} lines are very weak at about
10,000~K we did not try to determine helium abundances for the cool
stars. We note, however, that the observed \ion{He}{i} lines in these
cool stars agree well with those predicted for solar helium abundance
and thus with the results of Behr et al. (\cite{beco99},
\cite{beco00}) from high resolution spectroscopy.  Clearly the stable
stellar atmosphere required for radiative levitation also permits the
gravitational settling of helium. 

The mean magnesium abundance for the stars cooler than 12,000~K is
[Mg/H] = $-$1.4 and for the hotter stars it is $-$1.2. So the
magnesium abundance does not change significantly during the onset of
diffusion. The same result was also found by Moehler et
al. (\cite{mosw00}) for NGC~6752 and Behr et al. (\cite{beco99},
\cite{beco00}) for M~13 and M~15.

Rey et al. (\cite{reyo01}) have suggested that radiative levitation
and helium diffusion may be responsible for the formation of a blue
tail in M~13. However, we find that the HB stars in M~3 show the same
abundance pattern with respect to effective temperature as the stars
in M~13 and NGC~6752. The fact that M~3 contains stars affected by
radiative levitation
but does not have a blue tail provides strong evidence against
the Rey et al. (\cite{reyo01}) suggestion.

\begin{figure}
\vspace{11.5cm}
%\special{psfile=m3_m13_tg1.ps hscale=42 vscale=42 hoffset=-0 voffset=-5
\includegraphics{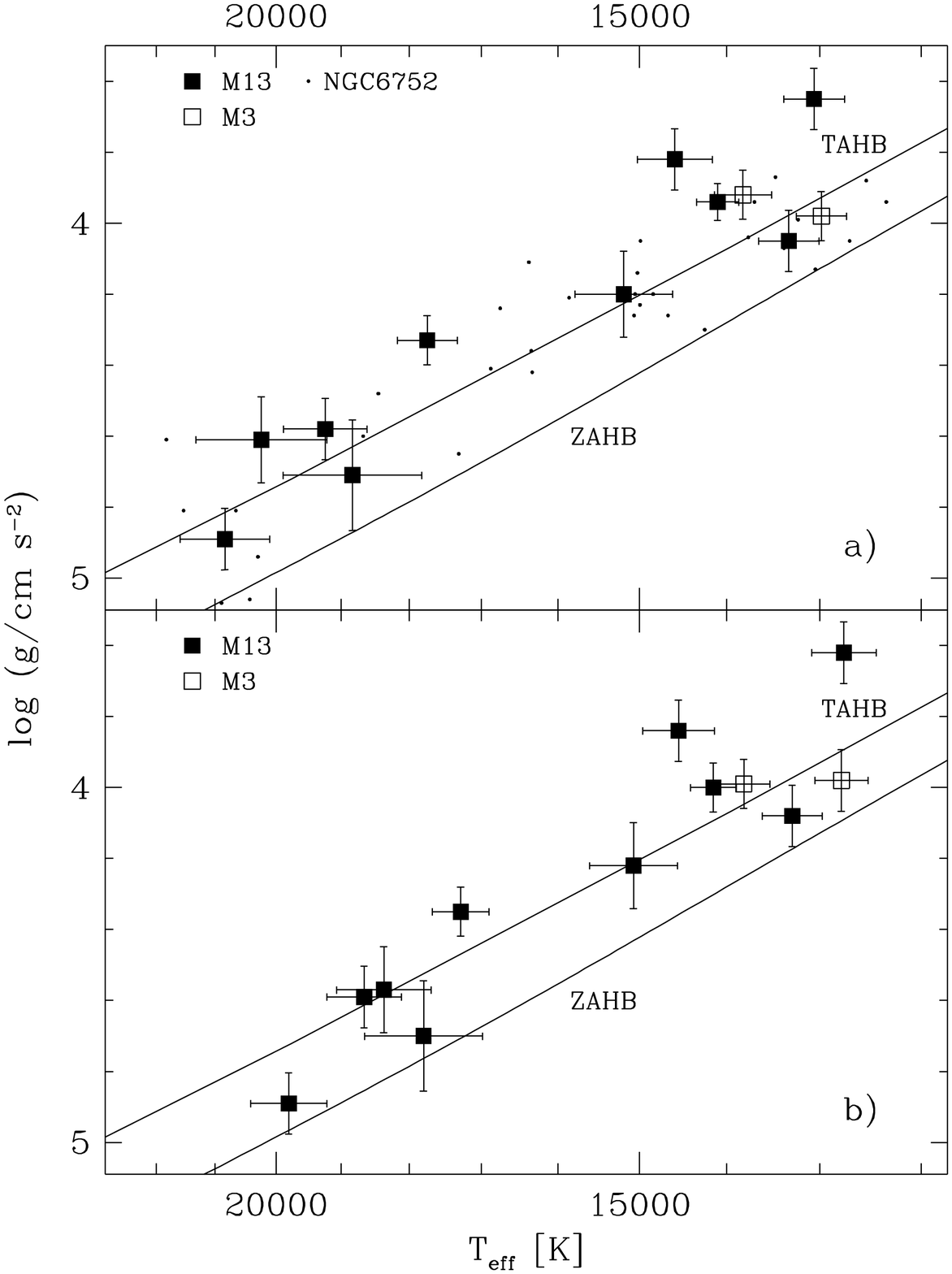}
\caption[]{
Temperatures and gravities of the programme stars that show evidence
for iron enrichment. The solid lines mark the zero-age (ZAHB) and terminal-age
(TAHB) loci of canonical HB tracks for [M/H] = $-$1.54. These loci
define the region within which the HB models spend 99\% of their HB
lifetime.\\ 
{\bf a)} Here we adopted a 
super-solar metallicity ([M/H] = $+$0.5) for the model
atmospheres (see Sect.~\ref{ssec_iron} for details), but did not
include metal lines in the theoretical spectra. 
For comparison we also show the results for blue tail stars in NGC~6752
from Moehler et al. (\cite{mosw00}, obtained with metal-rich model spectra
without metal lines).\\
{\bf b)} These are the results obtained with metal-rich model
spectra that include metal lines.
\label{m3_m13_tg1}} 
\end{figure}

   The stellar parameters derived for the stars hotter than 12,000 K
from the metal-rich model atmospheres are plotted in the upper panel
of Fig.~\ref{m3_m13_tg1} and compared to the canonical HB locus for a
helium abundance $Y$ of 0.23 and a scaled-solar metallicity [M/H] of
$-$1.54.  As already noted by Moehler et al. (\cite{mosw00}), using
scaled-solar metal-rich model atmospheres moves the stars hotter than
12,000 K closer to the canonical ZAHB, substantially reducing the low
gravity offset seen in Fig.~\ref{m3_m13_tg0}.  However, it is obvious that
discrepancies still exist between atmospheric parameters derived from
observations and the predictions of canonical HB theory.

The model spectra calculated from the metal-rich model atmospheres and
used for the results in the upper panel of Fig.~\ref{m3_m13_tg1} did
not include metal lines.  In order to determine if the strong metal
lines that show up at the onset of diffusion might influence our
results, we calculated a new grid of metal-rich model spectra, which
include metal lines, for temperatures up to 23,000 K, and then used
these spectra to fit all stars with effective temperatures above
12,000~K.  The results are plotted in Fig.~\ref{m3_m13_tg1} (lower panel)
and listed in Table~\ref{m3_m13_par_p05} (Cols. 6--8). While the
differences for stars cooler than 16000~K are small, the hotter stars
become considerably cooler when fitted with theoretical spectra that
include metal lines. This shift in temperature yields consistency of
the hottest stars with canonical HB evolution tracks. One should
remember, however, that these spectra were computed using scaled-solar
abundances. It is very unlikely that diffusion will produce an overall
enrichment of elements -- highly non-solar abundance ratios are much
more probable. So these results, while promising, should be taken with
a grain of salt.

Also, as can be seen from Fig.~\ref{m3_m13_mass}, the masses of the
stars obtained from their values of \teff\ and log g in
Tables~\ref{m3_m13_par_m15a} (Cols. 7 and 8) and \ref{m3_m13_par_p05}
(Cols. 6 and 7) tend to be too low.  Comparing the derived masses to
those expected for a star on the ZAHB at the same temperature yields a
mean mass ratio of 0.9$\pm$0.2 for the 5 stars below 12,000~K
(analysed with metal-poor model spectra including metal lines)
and 0.8$\pm$0.2 for the 12 hotter stars when analysed with
metal-rich model spectra including metal lines. We used distance
moduli of $(m-M)_V$ = \magpt{14}{51} (M~13) and \magpt{15}{02} (M~3)
to derive these masses.
           
We conclude that both the gravities and masses of the HB stars between
12,000~K and 16,000~K remain about 0.15--0.2 dex too low compared to
canonical predictions even when the stars are analyzed with metal-rich
model spectra which include metal lines. This remaining offset may
reflect a systematic error caused by our use of metal-rich model
atmospheres with scaled-solar abundances. The results of Behr et al.
(\cite{beco99}, \cite{beco00}) have demonstrated that radiative
levitation can lead to severely nonsolar abundance ratios in blue HB
stars.  Quite possibly, the use of scaled-solar metal-rich model
atmospheres may not sufficiently approximate the actual atmospheric
abundances in these stars. In addition, at such a high ([Fe/H] =
$+$0.5) metal abundance, the model atmospheres are not well tested,
and are more sensitive to inadequacies in the opacity distribution
function than are models at lower abundances. Also a stratification of
the stellar atmosphere influences the line profiles and our model
atmospheres assume homogeneous atmospheres. Another possibility,
suggested by Vink \& Cassisi (\cite{vica02}), is that an enhanced
stellar wind in the radiatively levitated HB stars may alter the wings
of the Balmer line profiles, leading to an underestimate of the
surface gravity.  While it remains to be seen if radiative levitation
can be effective at the high mass loss rates obtained by Vink \&
Cassisi (\cite{vica02}), their results do raise potential concerns
about the use of hydrostatic model atmospheres.

\section{Effects of varying helium abundance\label{sec-he}}
              
As mentioned in Sect.~\ref{sec_intro}, M~13 shows strong abundance
variations along its red giant branch, which might be attributed to
either deep mixing extending into the hydrogen-burning shell (helium
mixing) or pollution of the stars with the ejecta from an earlier
generation of AGB stars (helium pollution). Both processes would
increase the envelope helium abundance of the RGB stars with
important consequences for the subsequent HB evolution. Here we
discuss the potential impact of these effects on the properties of the
blue HB stars in M~13.

\subsection{Helium mixing\label{ssec-hemix}}

\begin{figure}
\vspace{11.5cm}
%\special{psfile=m3_m13_tg4.ps hscale=42 vscale=42 hoffset=-0 voffset=-5
\includegraphics{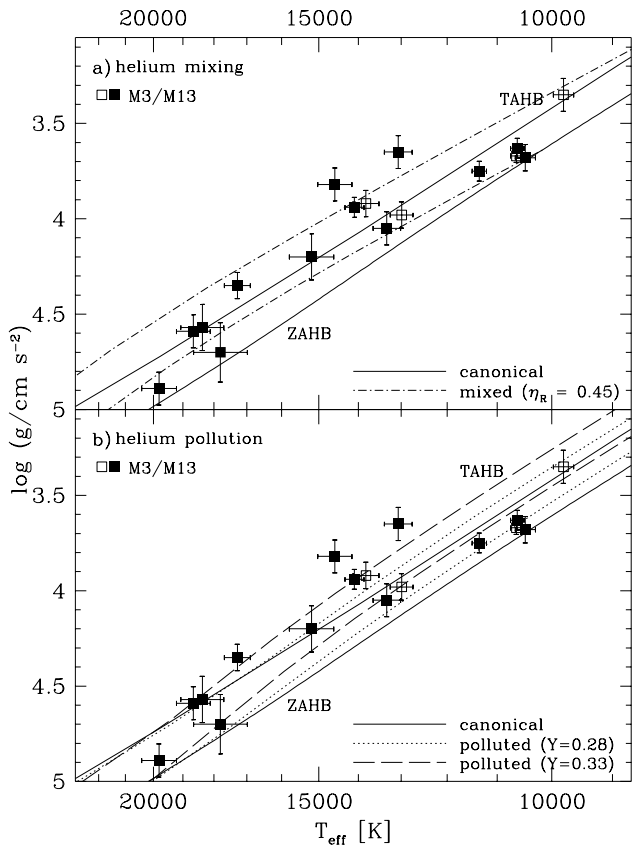}
\caption[]{ Temperatures and gravities of the programme stars in M~3 and
M~13 (from metal-rich model atmospheres for stars hotter than 12,000~K
and metal-poor model atmospheres for cooler stars, in both cases
including metal lines) compared to non-canonical evolutionary models.
The zero-age and terminal-age HB predicted by
canonical models for $Y$ = 0.23 and [M/H] = -1.54 are plotted as solid
lines.\\  {\bf a)} The zero-age and terminal-age HB for [M/H] = $-$1.54
computed with {\em helium mixing} for a Reimers mass-loss parameter
$\eta_R = 0.45$ (see Sect.~\ref{ssec-hemix} for details).\\  {\bf b)}
The zero-age and terminal-age HB computed with {\em helium pollution}
for helium abundances of $Y = 0.28$ and $Y = 0.33$ (see
Sect.~\ref{ssec-hepoll} for details).
%For more details see Fig.~\ref{m3_m13_tg0}.
\label{m3_m13_tg4}} 
\end{figure}

To illustrate the effects of helium mixing,
we will use a set of mixed sequences computed by Moehler et al. (\cite{mosw00})
for a main-sequence helium abundance $Y$ of 0.23 and a scaled-solar
metallicity [M/H] of $-$1.54. These sequences were evolved
up the RGB for different amounts of helium mixing using the
approach of Sweigart (\cite{swei97a}). Mass loss was included
according to the Reimers formulation with the mass-loss parameter $\eta_R$
set equal to 0.45. The evolution was then followed through the helium flash
to the end of the HB phase using standard
techniques. The HB
locus of these helium-mixed sequences in the \logg\ - \teff\
plane is shown in the top panel of
Fig.~\ref{m3_m13_tg4}.

Since mixing increases the RGB mass loss due to a brighter RGB tip luminosity,
a mixed model will arrive on the HB at a higher effective temperature
than the corresponding canonical model. At the same time mixing increases the
envelope helium abundance in the HB model, which leads
to a higher energy output of the hydrogen-burning shell and hence to
a brighter surface luminosity (Sweigart \& Gross \cite{swgr76}).

The net effect is to shift the mixed locus in Fig.~\ref{m3_m13_tg4}
towards lower gravities with increasing effective temperature relative to the
canonical locus, until a maximum offset is reached for 15,000~K $<$
\teff\ $<$ 20,000~K. At higher temperatures the mixed locus shifts
back towards the canonical locus, as the luminosity of the 
hydrogen-burning shell
declines due to the decreasing envelope mass. The predicted locus
along the extreme HB (EHB) does not depend strongly on the extent of
the mixing, since the luminosities and gravities of the EHB stars are
primarily determined by the mass of the helium core, which is nearly
the same for the mixed and canonical models. Models at the cool end of
the mixed locus will also differ little from the canonical models,
since the cool HB stars will have undergone little mixing on the
RGB.  This explains why the mixed and canonical loci converge at lower
temperatures in Fig.~\ref{m3_m13_tg4}. 
The size of the offset between the mixed and canonical
loci depends on the assumed value of $\eta_R$, becoming larger as
$\eta_R$ decreases (see Moehler et al. \cite{mosw00}).

\subsection{Helium pollution\label{ssec-hepoll}}
The ejecta of AGB stars which have undergone hot bottom burning should
be enriched in helium (Ventura et al. \cite{veda01}). Thus the
pollution of either pre-stellar material or existing low-mass stars by
this material will increase either the overall helium abundance or at
least that in the envelopes. D'Antona et al. (\cite{daca02}) have
studied the evolution of stars that form out of polluted material in
which the helium abundance was enriched from a canonical value of $Y =
0.23-0.24$ to $Y=0.29$. They found that while the effects on the
pre-HB evolution are quite small, the HB morphology itself is strongly
affected by helium pollution. The main effect is to shift the HB stars
to hotter temperatures due to smaller turnoff masses for a given
cluster age and to increased mass loss on the RGB.  Polluted stars
should thus be found primarily among the hottest HB stars.  Indeed,
D'Antona et al. (\cite{daca02}) suggest that helium pollution is
responsible for the extended blue tails found in such clusters as M~13
and NGC~6752.
           
To investigate the effects of helium pollution,
we computed two additional sets of HB sequences for [M/H] = $-$1.54
in which the helium abundance was increased by 0.05 and 0.10 above
our canonical value of $Y = 0.23$.  The HB loci defined by these
helium-polluted sequences are plotted in the bottom panel of
Fig.~\ref{m3_m13_tg4}.  The gravity offset between these loci
and the canonical locus is primarily due to the luminosity difference
between the helium-polluted and canonical models.  

HB stars derive
their energy from two sources: helium burning in the core and
hydrogen burning in a shell.  Since an increase in the initial
helium abundance leads to a decrease in the helium-core mass
(Sweigart \& Gross \cite{swgr78}), the energy output of the helium
core will be lower in a helium-polluted model than in a canonical
model.  Along the EHB, where hydrogen-shell burning is unimportant,
a helium-polluted star will therefore be fainter and have a higher gravity
than a canonical star.  The opposite is true at cooler temperatures
where the hydrogen-burning shell is a major energy source.  The
higher envelope helium abundance of a helium-polluted star
then leads to a brighter surface luminosity and thus to a lower
gravity.  At the transition between the blue HB and the EHB
(\teff\ $\approx$ 20,000~K), the gravities of the canonical and
helium-polluted models are virtually identical. This
is the point where the higher hydrogen-burning luminosity of
a helium-polluted star offsets the lower helium-burning
luminosity.
             
One aspect of the helium-polluted loci in Fig.~\ref{m3_m13_tg4}
requires further comment.  According to the helium-pollution
scenario outlined by D'Antona et al. (\cite{daca02}), the stars at
the cool end of an M~13-like HB, i.e., those near the top of
the blue tail, would not be helium polluted.  Thus in an actual cluster
one would expect the cooler HB stars to lie within the canonical
locus.  However, as one goes to higher temperatures along
the blue tail, the fraction of the HB stars that are helium
polluted would increase, and thus the locus predicted by
the D'Antona et al. scenario would shift away from the
canonical locus towards the lower gravities of the helium-polluted
loci in Fig.~\ref{m3_m13_tg4}.  This shift to lower
gravities is not seen in Fig.~\ref{m3_m13_tg4}, because the
helium-polluted loci in this figure assume that all of the
HB stars including the cooler stars are helium polluted.

\begin{figure}
\vspace{11.5cm}
%\special{psfile=m3_m13_mass.ps hscale=42 vscale=42 hoffset=-0 voffset=-5
\includegraphics{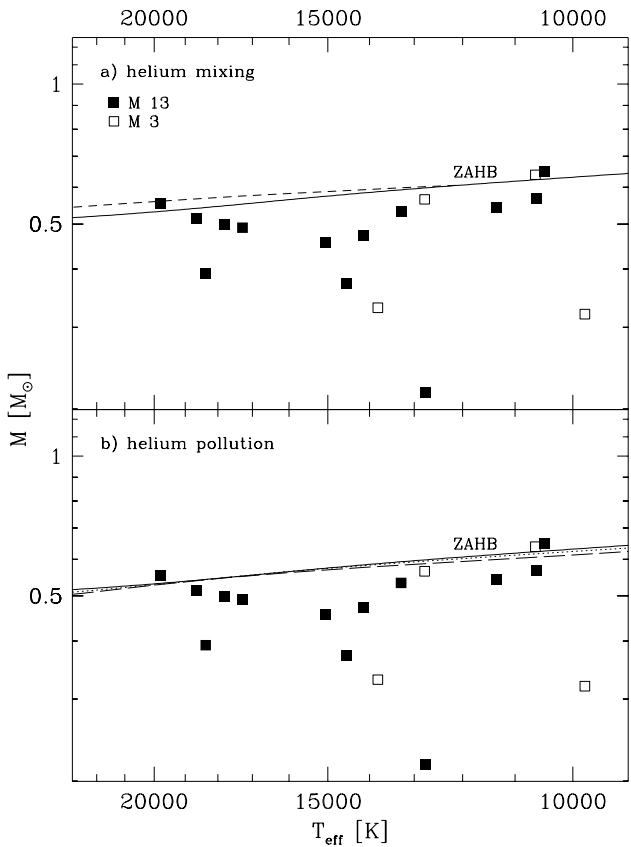}
\caption[]{Temperatures and masses of the programme stars in M~3 and
M~13 (from metal-rich model atmospheres for stars hotter than 12,000~K
and metal-poor model atmospheres for cooler stars; in both cases metal
lines were included in the theoretical spectra) compared to
evolutionary tracks.\\ 
{\bf a)} The solid line marks the
canonical ZAHB and the mixed ZAHB is given by the short-dashed line (see
Sect.~\ref{ssec-hemix} for details).\\  {\bf b)} Again the solid line marks 
the canonical
ZAHB and the polluted ZAHBs are given by the dotted (Y=0.28) and 
long-dashed line (Y=0.33), respectively (see Sect.~\ref{ssec-hepoll} for
details).
\label{m3_m13_mass}} 
\end{figure}

\subsection{Comparison with observations}
In Figs.~\ref{m3_m13_tg4} and \ref{m3_m13_mass} we compare the
effective temperatures, surface gravities and masses derived from
metal-rich (\teff\ $>$12,000~K) and metal-poor (\teff\ $<$12,000~K)
model atmospheres to the scenarios described above. While the
helium-mixed models provide a better description of the temperatures and
gravities derived for the stars between 12,000~K and 16,000~K
than the canonical models, they do
not resolve the problem with the low observed masses.
The same is also true for the helium-polluted models, where we again
find that the observed masses
are systematically too low.  The fact that neither the helium-mixed
nor helium-polluted models can account for the low masses raises doubt
about the reality of any gravity and/or temperature discrepancies between
the programme stars in Fig.~\ref{m3_m13_tg4} and canonical
predictions. Without a better understanding of possible systematic
errors, it is not possible to argue for or against the above noncanonical
scenarios, given the present data.

\section{Conclusions}

We have obtained low-resolution (3.4\AA) spectroscopy of 22 hot HB candidates
in M13, and four in M3, in order to derive atmospheric parameters (effective
temperatures and surface gravities) as well as abundances of helium, magnesium,
and iron.  One star (G43) in M13 turned out to be a UV-bright star, while the
remaining targets appear to be bona-fide HB stars.  For the stars between
10,000~K and 12,000~K, the atmospheric parameters derived from fitting the
Balmer lines and from Str\"omgren photometry are in good agreement.
For the stars cooler than 10,000~K, there is a discrepancy between  parameters
derived from Str\"omgren photometry and those derived from  fitting the Balmer
lines.    We suspect that there are problems in the Balmer line fitting  
because the derived temperatures show a gap between about  9000~K to 10,000~K,
and because the derived iron abundances show a wide scatter.   However,
despite numerous tests, we were unable to determine the cause of the 
disagreement in the derived parameters.

For stars hotter than 12,000~K in both clusters, we find evidence for  helium
depletion and a large iron enrichment, consistent with the results on M13 by
Behr et al. (\cite{beco99}) and on NGC~6752 by Moehler et al. (\cite{mosw00}).
The similar temperatures for the onset of radiative levitation in M3, M13, and
NGC~6752 suggest that this phenomenon is unrelated to the HB morphology.
Instead, the onset of radiative levitation may be connected to the
disappearance of the surface convection zone, as suggested by Sweigart 
(\cite{swei02}).

We compare our temperature and gravity results with the predictions of
the helium mixing scenario of Sweigart (\cite{swei97b}) and the helium
pollution scenario of D'Antona et al.  (\cite{daca02}).  These
scenarios are attractive because they can explain the origins of the
HB blue tail in globular clusters such as M13 without requiring a
fine-tuning of the mass loss, and because they can relate the blue
tail to the observed abundance anomalies on the RGB.  Both scenarios
predict lower gravities (and larger luminosities) for stars near
15,000 K.  For stars cooler than 12,000 K, we find the observed
gravities in agreement with canonical models.  This result is
consistent with the work of Caloi (\cite{calo01}), who compared the HB
luminosities of M3 and M13 to conclude that there was no evidence for
a substantial surface helium abundance increase for HB stars near the
temperature of the RR Lyrae stars.

For stars hotter than the onset of radiative levitation at 12,000 K,
we fit the Balmer lines using metal-rich ([Fe/H] = +0.5) model
atmospheres to derive the temperature and gravity.  Although the use
of metal-rich atmospheres reduces the discrepancy with canonical
models, we still find an offset to lower gravity of about 0.2 dex 
for stars between 12,000~K and 16,000~K, similar to what was seen at
temperatures hotter than 15,000~K in NGC~6752 by Moehler et
al. (\cite{mosw00}). However, there are a couple of reasons why this
result should be viewed with caution.  First, the derived masses for
these stars appear to be too low compared to either the canonical or the
helium-enriched scenarios.  Second, the metal-enriched atmospheres
used to derive these parameters are not well-tested, and do not take
into account the strongly non-solar abundance ratios.  Future
high-resolution spectroscopy could help reduce these possible
systematic errors in gravity, by allowing the atmospheric parameters
and abundances to be determined iteratively, using model atmospheres
with non-solar abundance ratios.

\begin{acknowledgements} We want to thank the staff of the Calar Alto
observatory for their support and B. Behr for making his abundance
measurements available to us. We are very grateful to F. Castelli and
M. Lemke for their help with the model atmospheres for the cool HB
stars. Thanks go also to our referee R.C. Peterson, whose comments
improved this paper considerably. This work was supported by the DLR
under grant 50\,OR\,96029-ZA.
\end{acknowledgements}

\end{document}